\documentclass[conference]{IEEEtran}

\usepackage{cite}
\usepackage{amsmath,amssymb,amsfonts}
\usepackage{booktabs}
\usepackage{multirow}
\usepackage{graphicx}
\usepackage[colorlinks=true,allcolors=blue]{hyperref}

\newcommand{\PUE}{\chi}
\newcommand{\Iset}{\mathcal{I}}
\newcommand{\Jset}{\mathcal{J}}
\newcommand{\Tset}{\mathcal{T}}
\newcommand{\Lset}{\mathcal{L}}
\newcommand{\Kset}{\mathcal{K}}
\newcommand{\Nset}{\mathcal{N}}
\newcommand{\Uset}{\mathcal{U}}
\newcommand{\Sset}{\mathcal{S}}

\makeatletter
\long\def\@makecaption#1#2{%
\ifx\@captype\@IEEEtablestring%
\footnotesize\bgroup\par\centering\@IEEEtabletopskipstrut{\normalfont\footnotesize #1}\\{\normalfont\footnotesize\scshape #2}\par\addvspace{0.5\baselineskip}\egroup%
\@IEEEtablecaptionsepspace
\else
\@IEEEfigurecaptionsepspace
\setbox\@tempboxa\hbox{\normalfont\footnotesize {#1.}\nobreakspace\nobreakspace #2}%
\ifdim \wd\@tempboxa >\hsize%
\setbox\@tempboxa\hbox{\normalfont\footnotesize {#1.}\nobreakspace\nobreakspace}%
\parbox[t]{\hsize}{\normalfont\footnotesize\noindent\unhbox\@tempboxa#2}%
\else%
\hbox to\hsize{\normalfont\footnotesize\box\@tempboxa\hfil}%
\fi\fi}
\makeatother

\begin{document}
\bstctlcite{IEEEexample:BSTcontrol}% "et al." beyond three authors (refs.bib)

\title{\vspace{-0.70\baselineskip}\fontsize{23}{25}\selectfont AI Inference-Aware Optimal Power Flow:\\
Converting Token Requirements into Grid Flexibility}

\author{\IEEEauthorblockN{Qianyi Ma, Dawei Qiu}
\IEEEauthorblockA{\textit{School of Electrical and Electronic Engineering}\\
\textit{Nanyang Technological University}\\
Singapore 639798, Singapore\\
qianyi.ma@ntu.edu.sg, dawei.qiu@ntu.edu.sg}
\and
\IEEEauthorblockN{Xiao Liu}
\IEEEauthorblockA{\textit{College of Engineering}\\
\textit{Cornell University}\\
Ithaca, NY 14853, USA\\
xl2372@cornell.edu}}

% slightly tighter gap between the author block and the abstract (Qianyi, 2026-11-07)
\IEEEaftertitletext{\vspace{-2.80\baselineskip}}
\maketitle

%==============================================================
\begin{abstract}
The rapid growth, temporal variability, and geographic concentration of electricity demand from artificial intelligence (AI) inference can aggravate transmission congestion near data centers (DCs). The spatial and temporal flexibility of inference workloads offers an opportunity to alleviate this congestion: requests can be routed across DCs or deferred within users’ quality-of-service (QoS) requirements. However, inference routing primarily considers graphics processing unit (GPU) availability, leaving this flexibility underused in grid operation.
Using GPU measurements, this paper models inference latency and DC power consumption to quantify the flexibility available under users’ time-to-first-token (TTFT) and inter-token latency (ITL) requirements. These models are then incorporated into an AI inference-aware optimal power flow (OPF) that coordinates inference routing and power system operation among DC operators, the transmission system operator, and inference users. By linking user requests, token generation at DCs, and electricity delivery through the transmission network, the formulation schedules inference workloads subject to both users’ QoS requirements and transmission constraints. 
Simulations on the PJM system show that, compared with computing-based routing and with electricity price-based routing, coordinating inference routing with network operation enables growing DC electricity demand deliverable to be met at lower cost.
\end{abstract}

\begin{IEEEkeywords}
AI inference, data centers, demand flexibility, optimal power flow, quality of service, tokens.
\end{IEEEkeywords}
\vspace{-0.28em}

%======================= Section I =====================
\section{Introduction}
\label{sec:intro}
Artificial intelligence (AI) inference accounts for an estimated 80–90\% of AI computing demand \cite{chung2026openg2g}, and its continued growth is expected to increase electricity consumption in data centers (DCs) \cite{epri2024powering}. Inference load may be concentrated at a small number of transmission buses and fluctuate with user requests \cite{wang2025burstgpt}. This concentration and variability can worsen local transmission congestion and limit the additional DC load that the network can supply.

At the same time, inference offers spatial and temporal flexibility that could help relieve this congestion. Each inference request generates a sequence of tokens, the basic units of model output. Users’ quality-of-service (QoS) requirements include limits on time-to-first-token (TTFT), the delay from request submission to the first output token, and inter-token latency (ITL), the interval between successive output tokens. Requests can be routed among geographically distributed DCs while respecting these latency limits. Users may also permit requests to be completed within a specified deferral window \cite{openai_batch_api}. A field trial has demonstrated that DC electricity use can be reduced during system peaks while maintaining workload-level QoS \cite{colangelo2026ai}.

In current practice, inference routing primarily considers computing resources and communication latency. Cloud platforms direct requests to DCs with available graphics processing units (GPUs) and low communication latency \cite{azure_deployment_types,aws_bedrock_cris}, while deferrable requests are scheduled to use otherwise idle GPUs \cite{sun2025hygen}. These mechanisms do not explicitly account for transmission constraints, leaving inference flexibility underused in power system operation.

Related research has approached inference flexibility from two perspectives. From the \textit{computing} perspective, studies have explored serving requests across regions to reduce GPU rental costs \cite{mao2025skyserve}, adapting inference operation to electricity prices \cite{stojkovic2025dynamollm,cheng2026learning}, and pricing inference services according to latency requirements \cite{sun2025energyaware}. From the \textit{power system} perspective, DCs have been modeled as responsive electricity loads \cite{du2026tokens} or spatially shiftable demand for emissions reduction \cite{han2026dispatchable,wan2026grid}, or participants in demand response programs \cite{spatialFlexibilityLFM,gameBasedCDL}.

Each perspective captures only part of the interaction among user requests, DC computing, and the power grid. Inference serving decisions determine the location, timing, and magnitude of DC electricity demand. However, computing-oriented studies often overlook how these decisions affect transmission congestion, leaving available flexibility underused in grid operation. Grid-oriented studies, in turn, may overestimate this flexibility by neglecting user QoS requirements or DC computing constraints. Two questions therefore remain: how much DC flexibility the grid can actually use while satisfying these requirements and constraints, and how inference serving can be coordinated with network operation.

To answer these questions, the main contributions of this paper are as follows:
\begin{enumerate}[\setlength{\IEEElabelindent}{0pt}\setlength{\labelsep}{0.3em}\IEEEsetlabelwidth{3)}\setlength{\topsep}{0pt}\setlength{\itemsep}{0pt}\setlength{\parsep}{0pt}]
\item An inference flexibility model is developed using GPU measurements. It links token generation to DC power consumption and quantifies the flexibility available to the grid under users’ TTFT limits, ITL limits, and request deferral windows.
\item An AI inference-aware OPF is proposed to coordinate inference routing and power system operation among the transmission system operator, DC operators, and inference users. It links user requests and token generation at DCs with electricity delivery through the transmission network under QoS, computing, and transmission constraints.
\item Case studies on the PJM system show that the proposed OPF accommodates growing inference demand at lower operating cost than computing-based and electricity price-based routing while respecting existing transmission limits.
\end{enumerate}
\vspace{-0.00em}

%=============== Section II ========================
\section{AI Inference-Aware OPF Framework}
\label{sec:framework}

\begin{figure}[!t]
    \vspace{-0.68em}
    \centering
    \includegraphics[width=\columnwidth]{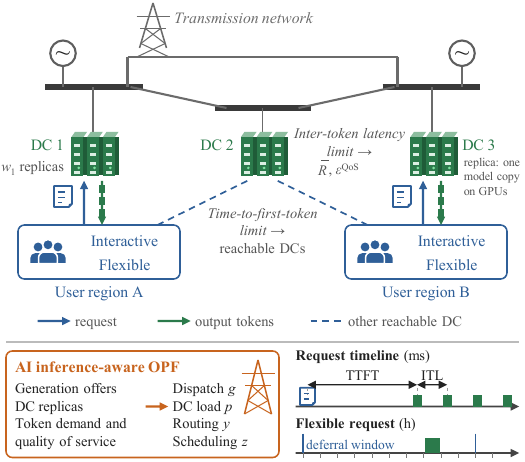}
    \vspace{-1.68em}
    \caption{Overview of the AI inference-aware OPF.}
    \label{fig:framework}
\end{figure}

Fig.~\ref{fig:framework} illustrates the coordinated operation among inference platforms, DC operators, and the transmission system operator. DC operators provide information on their installed model replicas and the measured relationship among replica throughput, latency, and GPU power consumption. Here, a replica denotes one copy of an AI model deployed on a specified group of GPUs. Inference platforms aggregate user requests by geographic region and QoS class, and provide the corresponding token demand, TTFT and ITL limits, and permissible deferral windows. The proposed AI inference-aware OPF uses these data as inputs together with generation offers and transmission network constraints.

These inputs constrain where and when inference requests can be served and how many tokens each active replica can produce. Specifically, the TTFT limit restricts which DCs can serve requests from each user region. The ITL requirement limits how many tokens an active replica can produce per hour. The deferral window specifies when demand may be served. Subject to these limits and transmission constraints, the inference-aware OPF selects where and when tokens are produced, how many replicas are activated, and how generators are dispatched.

These latency and deferral requirements vary across inference applications. The model therefore considers two request classes. Interactive requests, such as conversational inference, cannot be deferred and require a relatively fast token stream. Flexible requests, such as batch inference jobs, allow a slower token stream and may be deferred within a specified window. The OPF uses hourly intervals to represent deferral across hours. Inference platforms implement the resulting token allocations by routing and scheduling individual requests within each hour.
\vspace{-0.00em}

%================ Section III ======================
\section{AI Inference-Aware OPF Model}
\label{sec:model}
The system operator maximizes social welfare, i.e., the utility of the served inference demand minus the generation cost. Since all demand must be served, this utility is fixed, and the problem reduces to minimizing the total generation cost over the hours $t\in\Tset$, subject to the request constraints \eqref{eq:reach}--\eqref{eq:c-service}, the DC constraints \eqref{eq:c-group}--\eqref{eq:group-power}, and the network constraints \eqref{eq:c-bal}--\eqref{eq:gen}:
\vspace{-0.68em}
\begin{equation}
    \min_{z,y,r,x,g}\ \textstyle\sum_t\sum_k C_k(g_{k,t})
    \quad\text{s.t. }\eqref{eq:reach}\text{--}\eqref{eq:gen},
    \label{eq:obj}
\end{equation}
where $g_{k,t}$ is the output of generator $k$ and $C_k$ its cost function.

\subsection{Inference Requests Model}
\label{sec:requests}
Users in region $u\in\Uset$ experience approximately the same communication round-trip time (RTT) $L^{\rm RTT}_{u,i}$ to DC $i\in\Iset$. A QoS class $s\in\Sset$ specifies a TTFT limit $\bar L_s^{\rm TTFT}$, an ITL limit $\bar L_s^{\rm ITL}$, and a deferral window $\bar\delta_s$ in hours. Each region--class pair defines a request group $j=(u,s)\in\Jset$ with token demand $D_{j,\tau}$ arriving in hour $\tau$.

TTFT is approximated by RTT plus a fixed prefill time $\tau_i^{\rm pre}$ for processing the input prompt. The TTFT limit therefore determines where a request can be served, namely at the reachable DCs of group $j$:
\begin{equation}
    \Iset_j=\{i\in\Iset:\ L^{\rm RTT}_{u(j),i}+\tau_i^{\rm pre}\le\bar L^{\rm TTFT}_{s(j)}\}.
    \label{eq:reach}
\end{equation}
The deferral window, in turn, determines when a request can be served. Let $z_{j,\tau,t}$ be the demand of group $j$ arriving in hour $\tau$ and served in hour $t$. The demand arriving in hour $\tau$ must be fully served within its deferral window $[\tau,\tau+\bar\delta_{s(j)}]$:
\begin{equation}
    D_{j,\tau}-\textstyle\sum_{t=\tau}^{\tau+\bar\delta_{s(j)}}z_{j,\tau,t}=0.
    \label{eq:c-balance}
\end{equation}
For flexible requests, the deferral window applies to the delay before service begins, whereas the modeled TTFT and ITL limits apply once serving starts. Let $y_{ji,t}$ be the tokens of group $j$ served at DC $i$ in hour $t$. Whatever their arrival hour, the tokens served in hour $t$ are allocated among the reachable DCs:
\begin{equation}
    \textstyle\sum_{\tau:\,t\in[\tau,\tau+\bar\delta_{s(j)}]}z_{j,\tau,t}-\sum_{i\in\Iset_j}y_{ji,t}=0.
    \label{eq:c-service}
\end{equation}
Interactive requests set $\bar\delta_s=0$ and are served on arrival.

\subsection{Data Center Model}
\label{sec:dc}
\begin{figure}[!t]
\vspace{-1.28em}
    \centering
    \includegraphics[width=0.95\columnwidth]{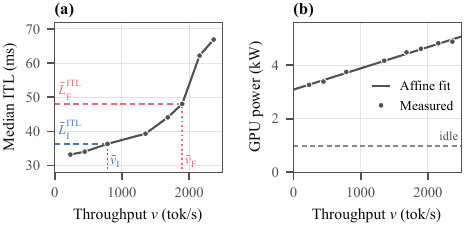}
    \vspace{-0.88em}
    \caption{Measured GPU serving curves versus throughput: (a) inter-token latency; (b) GPU power.}
    \label{fig:dc-model}
\end{figure}

Each DC $i$ hosts a fixed number $w_i$ replicas. In hour $t$, $x_{i,s,t}$ of replicas are activated for QoS class $s$, and the others remain idle. The variable $x_{i,s,t}$ is continuous, which allows replicas to be shared within an hour. A replica raises its throughput $v$, i.e., its token generation rate, by serving more requests concurrently, at the expense of a longer ITL. Fig.~\ref{fig:dc-model} captures this trade-off with an increasing ITL curve $L_i(v)$ and an affine GPU power fit $a_i+\beta_i v$. The ITL limit of class $s$ therefore caps the throughput of a replica at $\bar v_{i,s}$, defined by $L_i(\bar v_{i,s})=\bar L_s^{\rm ITL}$, so that a replica produces at most $\bar R_{i,s}=\kappa\bar v_{i,s}$ class-$s$ tokens in an hourly interval of $\kappa$ seconds. The class-$s$ token output $r_{i,s,t}$ of DC $i$ must equal the tokens routed to it. Since a request is completed within seconds, routing and production take place in the same hour:
\vspace{-0.38em}
\begin{equation}
    \textstyle\sum_{j:\,s(j)=s,\ i\in\Iset_j}y_{ji,t}-r_{i,s,t}=0.
    \label{eq:c-group}
\end{equation}
Since each active replica produces at most $\bar R_{i,s}$ tokens per hour, enough replicas must be activated to cover this output:
\vspace{-0.38em}
\begin{equation}
    r_{i,s,t}-\bar R_{i,s}x_{i,s,t}\le0,
    \label{eq:c-sla-cap}
\end{equation}
so that each class-$s$ token occupies a fraction $1/\bar R_{i,s}$ of a replica.
% Fig. 2 defined here so that it floats to the top of the first column of page 3 

Finally, the replicas activated for all classes cannot exceed those hosted by the DC:
\begin{equation}
    \textstyle\sum_s x_{i,s,t}-w_i\le0.
    \label{eq:c-replica}
\end{equation}
With fixed base load $p_i^{\rm base}$, idle replica power $p_i^{\rm idle}$, and power usage effectiveness $\PUE_i$, DC power is
\begin{equation}
    p_{i,t}=p_i^{\rm base}+\PUE_iw_ip_i^{\rm idle}
    +h_i\textstyle\sum_s x_{i,s,t}+e_i\sum_s r_{i,s,t},
    \label{eq:group-power}
\end{equation}
where $h_i=\PUE_i(a_i-p_i^{\rm idle})$ is the additional power of activating a replica and $e_i=\PUE_i\beta_i/\kappa$ is the marginal energy consumption per token. Thus, \eqref{eq:group-power} splits DC power consumption into a constant component comprising base load and idle consumption, an activation component proportional to the number of active replicas, and a token-production component proportional to token output. A tighter ITL limit can reduce the token capacity of each replica. Serving the same token demand may therefore require more active replicas, increasing the activation overhead and hence the average energy consumption per token.

\subsection{Transmission Network Constraints}
\label{sec:opf}
DC $i$ draws $p_{i,t}$ at bus $n(i)$ of a network with buses $n\in\Nset$, lines $\ell\in\Lset$, and generators $k\in\Kset$. Under the direct-current power-flow approximation, with other load $L^{\rm oth}_{n,t}$ and power transfer distribution factors $\Psi_{\ell n}$,
\begin{align}
    &\textstyle\sum_i p_{i,t}+\sum_n L^{\rm oth}_{n,t}-\sum_k g_{k,t}=0,
    \label{eq:c-bal}\\
    &-\bar f_\ell\le\textstyle\sum_n\Psi_{\ell n}\mathrm{inj}_{n,t}\le\bar f_\ell,
    \label{eq:c-line}\\
    &\mathrm{inj}_{n,t}=\textstyle\sum_{k:n(k)=n}g_{k,t}-L^{\rm oth}_{n,t}-\sum_{i:n(i)=n}p_{i,t},
    \label{eq:inj}\\
    &\underline G_k\le g_{k,t}\le\bar G_k,
    \label{eq:gen}
\end{align}
where $\bar f_\ell$ is the line-flow limit, $\underline G_k$ and $\bar G_k$ are generation limits, and $z,y,r,x\ge0$. Constraint \eqref{eq:c-bal} balances total generation with the DC load and the other load. Constraint \eqref{eq:c-line} keeps the line flows, obtained from the net bus injections \eqref{eq:inj} through the power transfer distribution factors, within their limits. Finally, constraint \eqref{eq:gen} bounds the generator outputs.

In summary, the decision variables are the generator outputs $g$, the deferral schedule $z$, the token routing $y$, and the DC token outputs $r$ and active replicas $x$. And the proposed OPF is a convex program.
\vspace{-0.28em}

%=============== Section IV ====================
\section{Case Studies}
\label{sec:case}
\begin{figure}[t]% [t], not [!t]: "!" would override \suppressfloats above
\vspace{-0.48em}
    \centering
    \includegraphics[width=2.8in]{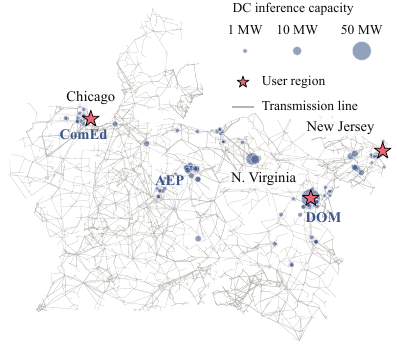}
    \vspace{-1.08em}
    \caption{PJM test system with DCs and user regions.}
    \label{fig:setting}
\end{figure}

\suppressfloats[t]% Fig. 3 and Table I must not float above this heading 
Case studies use the 7,830-bus PJM network \cite{britto2026powergrid}. Solar profiles are derived from the National Solar Radiation Database \cite{sengupta2018nsrdb}, wind profiles from bias-corrected High-Resolution Rapid Refresh data \cite{buster2024bchrrr}, and load profiles from EIA-930 \cite{eia930}. A total of 333 DCs are identified from public inventories, including the IM3 atlas \cite{mongird2026im3}, and mapped to 138 buses. Fig.~\ref{fig:setting} shows the resulting test system. The main DC clusters are in Northern Virginia (DOM), Ohio (AEP), and the Chicago area (ComEd).

At each DC, 10\% of capacity is allocated to inference. Each replica serves Llama-3.1-405B on eight H100 GPUs, with serving curves fitted to measurements from the ML.ENERGY Benchmark \cite{chung2025mlenergy}. Requests originate from three regions: Chicago, Northern Virginia, and New Jersey. RTTs are obtained from RIPE Atlas \cite{ripeatlas2015}, and request demand profiles from BurstGPT \cite{wang2025burstgpt}. The median ITL limits are 36.3~ms for interactive requests and 48.0~ms for flexible requests. Both classes have an 80-ms TTFT limit, with a 40-ms prefill time.

Weekly simulations cover July 10--16, 2023, with cyclic hour indexing so that deferral windows wrap around the end of the week. Single-hour analyses use the snapshot at 16:00 on July 15, 2023. All optimization models are solved using Gurobi 12.0.3.
\vspace{-0.28em}

\subsection{Superiority of the Inference-Aware OPF}
\label{sec:deliverability}

\begin{table}[t]% [t], not [!t]: "!" would override \suppressfloats above
\vspace{-0.68em}
\centering
\caption{Comparison of different inference routing schemes}
\vspace{-0.68em}
\label{tab:schemes}
\footnotesize
\setlength{\tabcolsep}{2.5pt}
% units on their own row so the table is narrower and \resizebox enlarges the font
\resizebox{\columnwidth}{!}{%
\begin{tabular}{@{}lccccccc@{}}
\toprule
Inference & \multicolumn{3}{c}{Infeasible hours} & \multicolumn{4}{c}{Inference-induced system cost}\\
\cmidrule(lr){2-4}\cmidrule(l){5-8}
energy & \multirow{2}{*}{Computing} & \multirow{2}{*}{Ex-ante LMP}& \multirow{2}{*}{Proposed} & \multirow{2}{*}{Computing} & \multirow{2}{*}{Ex-ante LMP}& \multicolumn{2}{c}{Proposed}\\
\cmidrule(l){7-8}
share & & & & & & w/o def. & 12-h def.\\
 & & & & ($\Delta$\%) & ($\Delta$\%) & ($\Delta$\%) & (M\$)\\
\midrule
0.3\% (base) & 0 & 0 & 0& $+$9.0 & $+$8.4 & $+$7.5 & 0.91\\
2\% & 0 & 0 & 0& $+$8.1 & $+$7.5 & $+$6.6 & 6.65\\
4\% & 0 & 0 & 0& $+$7.1 & $+$6.6 & $+$5.7 & 13.93\\
6\% & 20 & 10 & 0& -- & -- & $+$4.7 & 21.90\\
8\% & 83 & 16 & 0& -- & -- & $+$3.9 & 30.59\\
\bottomrule
\end{tabular}}
\end{table}

To evaluate the benefit of coordinating inference routing with network operation, the proposed OPF is compared with two routing benchmarks: 1) \textit{Computing-based routing} (prevailing industry practice) assigns each request to the nearest reachable DC, i.e., the one with the lowest RTT, among those with spare GPU capacity \cite{azure_deployment_types,aws_bedrock_cris}; 2) \textit{Ex-ante LMP-based routing} assigns each request to the reachable DC with the lowest electricity cost per token, evaluated using LMPs computed without inference load, and routes excess demand to the next cheapest reachable DC \cite{gameBasedCDL}. GPU capacity and token demand are scaled at the existing DC buses to obtain inference energy shares of 2--8\%.

Table~\ref{tab:schemes} reports the number of network-infeasible hours and the weekly system cost added by inference. Relative cost increases are calculated using the inference-aware OPF with a 12-h deferral window as the reference. At low inference shares, when both benchmarks remain feasible, computing-based routing yields the highest cost, followed by LMP-based routing. Computing-based routing minimizes latency without considering electricity prices. LMP-based routing uses prices and respects GPU capacity limits, but does not jointly consider the alternative DCs available to different user regions. It may therefore assign scarce low-cost capacity to requests with other affordable options, leaving requests with fewer options to be served at expensive DCs. The proposed OPF instead jointly optimizes routing and GPU allocation, reserving low-cost capacity to requests with fewer affordable alternatives and directing requests with more options elsewhere. This reduces reliance on expensive DCs and lowers the overall system cost.

At inference shares of 6\% and 8\%, both benchmark schemes become network-infeasible during weekday afternoons, whereas the proposed OPF remains feasible in every hour. Under computing-based routing, requests from Northern Virginia concentrate at nearby DCs in the DOM zone, where the additional load exceeds the zone's available import capability. LMP-based routing produces fewer infeasible hours but also concentrates demand at low-price DCs, since its routing prices do not reflect the additional congestion caused by inference load. The proposed OPF avoids these violations by routing part of the demand to DCs in Ohio and western Pennsylvania that remain reachable within the users' TTFT limits. Allowing deferral further reduces the system cost, although the percentage savings decline as the inference share increases. This is because the additional demand shifted to night-time hours raises LMPs in those hours, reducing the savings from off-peak operation.

\begin{figure}[!t]
\vspace{-0.48em}
    \centering
    \includegraphics[width=\columnwidth]{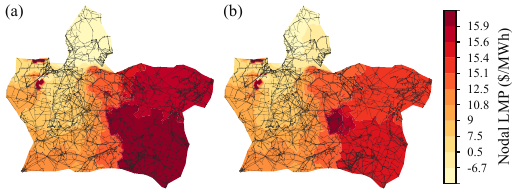}
    \vspace{-1.88em}
    \caption{Nodal LMPs at an 8\% inference share: (a) LMP-based routing; (b) proposed AI inference-aware OPF.}
    \label{fig:lmp-map}
\end{figure}

Fig.~\ref{fig:lmp-map} compares nodal LMPs under LMP-based routing and the proposed OPF for the single-hour snapshot at an 8\% inference energy share. The proposed OPF routes part of the inference demand to northern areas where exports of local generation are constrained, allowing more generation to be consumed locally. LMPs there rise from negative levels, while those in the high-price eastern areas decline. This reallocation narrows the spatial spread of LMPs, and congestion rent also falls without network reinforcement.

\subsection{Sensitivity Analysis of Inference Requirements}
\label{sec:latency-cost}

\begin{figure}[!t]
\vspace{-0.48em}
    \centering
    \includegraphics[width=0.93\columnwidth]{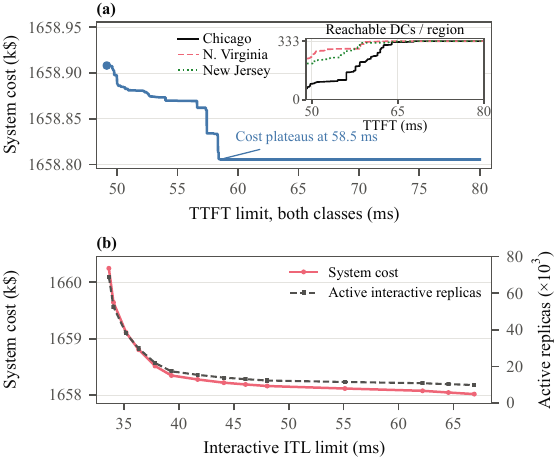}
    \vspace{-0.48em}
    \caption{System cost under varying (a) TTFT and (b) interactive ITL limits.}
    \label{fig:latency-cost}
\end{figure}

Fig.~\ref{fig:latency-cost} shows how system cost changes when the TTFT and ITL limits are relaxed separately, with all other parameters held fixed. In Fig.~\ref{fig:latency-cost}(a), relaxing the TTFT limit makes additional DCs reachable, reducing system cost in steps. Once the low-cost DCs are reachable from every user region, further relaxation no longer reduces cost. In Fig.~\ref{fig:latency-cost}(b), relaxing the ITL limit reduces both system cost and the number of active replicas serving interactive requests. A looser ITL limit allows each replica to serve more requests at once. The same demand can therefore be served with fewer active replicas, reducing activation power.

In summary, the two limits affect system cost through different mechanisms: relaxing TTFT expands the choice of serving locations, whereas relaxing ITL increases the token capacity of each replica and reduces the associated activation overhead. Over the tested ranges, system cost is more sensitive to the ITL limit than to the TTFT limit. The comparison highlights the value of inference-serving efficiency as a source of grid flexibility. 

\subsection{Temporal Flexibility of Inference}
\label{sec:deferral}

\begin{figure}[!t]
\vspace{-0.48em}
    \centering
    \includegraphics[width=0.93\columnwidth]{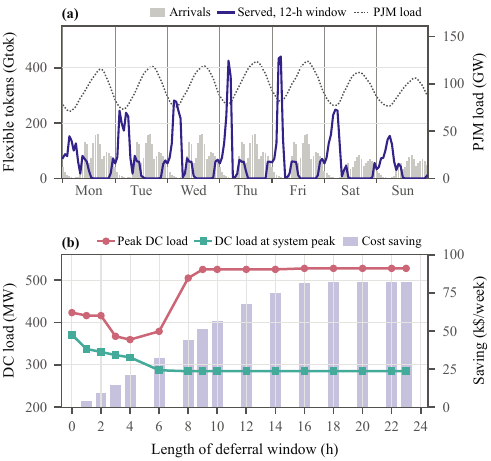}
    \vspace{-0.48em}
    \caption{Temporal flexibility of inference: (a) flexible-token service and PJM load under the 12-h deferral window; (b) DC load and cost saving versus deferral window.}
    \label{fig:dc-load}
\end{figure}

Fig.~\ref{fig:dc-load}(a) the arrivals and service of flexible tokens under a 12-h deferral window. Deferral shifts afternoon arrivals into the night, when LMPs are low and GPUs are idle, which lowers both the generation cost and the DC load at the daily PJM peak.
As the PJM night load rises on Thursday and Friday, fewer cheap hours remain, and the deferred demand concentrates into sharp night peaks.

Fig.~\ref{fig:dc-load}(b) shows how the length of the deferral window affects the weekly saving and the DC load. Short windows already shift the afternoon arrivals away from the system peak, so the DC load at that peak stops declining once the window reaches 6~h. Weekly savings continue to grow, however, as longer windows allow demand to be served during the cheapest night hours. With long windows, deferred demand concentrates into sufficiently few hours that peak DC load exceeds its level without deferral. Longer windows therefore provide additional cost savings but can increase night DC peaks, which remain subject to the transmission constraints enforced by the OPF.
\vspace{-0.48em}

%==============================================================
\section{Conclusion}
\vspace{-0.18em}
\label{sec:conclusion}
This paper proposed an AI inference-aware OPF that jointly determines inference routing, replica activation, and generator dispatch under users’ QoS requirements and transmission constraints. Case studies on the PJM system show that this coordination enables growing DC electricity demand to be met at lower operating cost on the existing transmission network. Users’ latency tolerance provides flexibility by expanding routing options and increasing the token capacity of each replica, thereby reducing activation overhead. Longer deferral windows yield further savings but can concentrate demand into higher night-time peaks at DC buses. By incorporating users’ latency and deferral limits as explicit optimization constraints, the framework provides a basis for inference platforms to offer grid-responsive services.
\vspace{-0.48em}

%==============================================================
\bibliographystyle{IEEEtran}
\bibliography{refs}

@IEEEtranBSTCTL{IEEEexample:BSTcontrol,
  CTLuse_forced_etal       = "yes",
  CTLmax_names_forced_etal = "3",
  CTLnames_show_etal       = "3"
}

@article{han2026dispatchable,
  author  = {Jianpei Han and Ershun Du and Bojun Du and Yaowang Li and Jingrong Guo and Ning Zhang and Chongqing Kang},
  title   = {Evaluating the Dispatchable Capacity of Cross-Regional Data Center Clusters Toward Power System Operation},
  journal = {IEEE Transactions on Smart Grid},
  year    = {2026},
  volume  = {17},
  number  = {2},
  pages   = {1180--1193},
  month   = mar,
  doi     = {10.1109/TSG.2025.3623107}
}

@misc{openai_batch_api,
  author       = {{OpenAI}},
  title        = {Batch {API}},
  howpublished = {OpenAI API documentation},
  note         = {Accessed Oct. 2026},
  url          = {https://developers.openai.com/api/docs/guides/batch}
}

@inproceedings{mao2025skyserve,
  author    = {Ziming Mao and Tian Xia and Zhanghao Wu and Wei-Lin Chiang and Tyler Griggs and Romil Bhardwaj and Zongheng Yang and Scott Shenker and Ion Stoica},
  title     = {{SkyServe}: Serving {AI} Models across Regions and Clouds with Spot Instances},
  booktitle = {Proc. 20th European Conf. Computer Systems (EuroSys)},
  year      = {2025},
  pages     = {159--175},
  doi       = {10.1145/3689031.3717459}
}

@inproceedings{stojkovic2025dynamollm,
  author    = {Jovan Stojkovic and Chaojie Zhang and {\'I}{\~n}igo Goiri and Josep Torrellas and Esha Choukse},
  title     = {{DynamoLLM}: Designing {LLM} Inference Clusters for Performance and Energy Efficiency},
  booktitle = {Proc. IEEE Int. Symp. High Performance Computer Architecture (HPCA)},
  year      = {2025},
  pages     = {1348--1362},
  doi       = {10.1109/HPCA61900.2025.00102}
}

@techreport{epri2024powering,
  author      = {{Electric Power Research Institute}},
  title       = {Powering Intelligence: Analyzing Artificial Intelligence and Data Center Energy Consumption},
  institution = {EPRI},
  address     = {Palo Alto, CA},
  number      = {3002028905},
  type        = {White Paper},
  month       = may,
  year        = {2024}
}

@inproceedings{sun2025hygen,
  author    = {Ting Sun and Penghan Wang and Fan Lai},
  title     = {{HyGen}: Efficient {LLM} Serving via Elastic Online-Offline Request Co-Location},
  booktitle = {Proc. Advances in Neural Information Processing Systems (NeurIPS)},
  year      = {2025}
}

@misc{aws_bedrock_cris,
  author       = {{Amazon Web Services}},
  title        = {Increase throughput with cross-Region inference},
  howpublished = {Amazon Bedrock User Guide},
  note         = {Accessed Oct. 2026},
  url          = {https://docs.aws.amazon.com/bedrock/latest/userguide/cross-region-inference.html}
}

@misc{azure_deployment_types,
  author       = {{Microsoft}},
  title        = {Azure {OpenAI} deployment types},
  howpublished = {Microsoft Learn},
  note         = {Accessed Oct. 2026},
  url          = {https://learn.microsoft.com/en-us/azure/ai-services/openai/how-to/deployment-types}
}

@article{britto2026powergrid,
  title   = {Building Power Grid Models from Open Data: A Complete Pipeline from OpenStreetMap to Optimal Power Flow},
  author  = {Britto, Andrea and Spina, Thiago and Yang, Weiwei and Fowers, Spencer and Zhang, Baosen and White, Chris},
  journal = {arXiv preprint arXiv:2605.04289},
  year    = {2026}
}

@misc{buster2024bchrrr,
  title        = {Bias Corrected NOAA HRRR Wind Resource Data for Grid Integration Applications},
  author       = {Buster, Grant and Pinchuk, Pavlo and Lavin, Luke and Benton, Brandon and Bodini, Nicola},
  year         = {2024},
  howpublished = {Open Energy Data Initiative, National Renewable Energy Laboratory},
  doi          = {10.25984/2480349}
}

@inproceedings{chung2025mlenergy,
  title     = {The {ML.ENERGY Benchmark}: Toward Automated Inference Energy Measurement and Optimization},
  author    = {Chung, Jae-Won and Ma, Jeff J. and Wu, Ruofan and Liu, Jiachen and Kweon, Oh Jun and Xia, Yuxuan and Wu, Zhiyu and Chowdhury, Mosharaf},
  booktitle = {NeurIPS Datasets and Benchmarks},
  year      = {2025}
}

@article{colangelo2026ai,
  title   = {{AI} data centres as grid-interactive assets},
  author  = {Colangelo, Philip and Coskun, Ayse K. and Megrue, Jack and Roberts, Ciaran and Sengupta, Shayan and Sivaram, Varun and Tiao, Ethan and Vijaykar, Aroon and Williams, Chris and Wilson, Daniel C. and Records, Brandon and MacFarland, Zack and Dreiling, Daniel and Morey, Nathan and Ratnayake, Anuja and Vairamohan, Baskar},
  journal = {Nature Energy},
  volume  = {11},
  pages   = {254--261},
  year    = {2026},
  doi     = {10.1038/s41560-025-01927-1}
}

@article{du2026tokens,
  title={From Tokens to Energy Flexibility: Quantization-Enabled Demand Response for Data Centers with {LLM} Inference Workloads},
  author={Du, Bojun and Fan, Xiaoyi and Du, Ershun and Chen, Long and Han, Jianpei and Hou, Qingchun and Zhang, Ning and Kang, Chongqing},
  journal={arXiv preprint arXiv:2606.18851},
  year={2026},
  doi={10.48550/arXiv.2606.18851}
}

@misc{eia930,
  author       = {{U.S. Energy Information Administration}},
  title        = {Form EIA-930: Hourly and Daily Balancing Authority Operations Report},
  howpublished = {Hourly Electric Grid Monitor},
  url          = {https://www.eia.gov/electricity/gridmonitor/}
}

@article{gameBasedCDL,
  title={Game-Based Optimization Method for Geo-Distributed Data Centers Under Customer Directrix Load Demand Response Mechanism},
  author={Wu, Jia-Kai and Liu, Zhi-Wei and Zhao, Yong and Li, Wenqu and Li, Yuanzheng},
  journal={IEEE Transactions on Smart Grid},
  volume={17},
  number={1},
  pages={283--296},
  year={2026},
  doi={10.1109/TSG.2025.3612048}
}

@misc{mongird2026im3,
  title   = {IM3 Open Source Data Center Atlas},
  author  = {Mongird, Kendall and Thurber, Travis and Vernon, Chris and Burleyson, Casey and Akdemir, Kerem Ziya and Rice, Jennie},
  year    = {2026},
  note    = {Version 2026.02.09, MSD-LIVE Data Repository},
  doi     = {10.57931/3017294}
}

@article{ripeatlas2015,
  title   = {{RIPE Atlas}: A Global Internet Measurement Network},
  author  = {{RIPE NCC Staff}},
  journal = {The Internet Protocol Journal},
  volume  = {18},
  number  = {3},
  pages   = {2--26},
  year    = {2015}
}

@article{sengupta2018nsrdb,
  title   = {The National Solar Radiation Data Base ({NSRDB})},
  author  = {Sengupta, Manajit and Xie, Yu and Lopez, Anthony and Habte, Aron and Maclaurin, Galen and Shelby, James},
  journal = {Renewable and Sustainable Energy Reviews},
  volume  = {89},
  pages   = {51--60},
  year    = {2018},
  doi     = {10.1016/j.rser.2018.03.003}
}

@article{spatialFlexibilityLFM,
  title={Spatial Flexibility Provision From Geographically Dispersed Data Centers Enabling Coordinated Operation of Multi-Local Flexibility Markets},
  author={Chen, Boyu and Che, Yanbo and Takci, Mehmet T{\"u}rker and Qadrdan, Meysam and Zhou, Yue},
  journal={IEEE Transactions on Smart Grid},
  volume={17},
  number={2},
  pages={1371--1381},
  year={2026},
  doi={10.1109/TSG.2025.3648617}
}

@inproceedings{wang2025burstgpt,
  title     = {{BurstGPT}: A Real-World Workload Dataset to Optimize LLM Serving Systems},
  author    = {Wang, Yuxin and Chen, Yuhan and Li, Zeyu and Kang, Xueze and Fang, Yuchu and Zhou, Yeju and Zheng, Yang and Tang, Zhenheng and He, Xin and Guo, Rui and Wang, Xin and Wang, Qiang and Zhou, Amelie Chi and Chu, Xiaowen},
  booktitle = {Proceedings of the 31st ACM SIGKDD Conference on Knowledge Discovery and Data Mining},
  year      = {2025},
  publisher = {ACM},
  doi       = {10.1145/3711896.3737413}
}

@inproceedings{sun2025energyaware,
  author    = {Yimeng Sun and Xinru Guan and Zhaohao Ding},
  title     = {Energy-aware Computing Pricing for Cloud Data Centers},
  booktitle = {Proc. IEEE Power \& Energy Society General Meeting (PESGM)},
  address   = {Austin, TX, USA},
  year      = {2025},
  doi       = {10.1109/PESGM52009.2025.11225529}
}

@inproceedings{cheng2026learning,
  author    = {Yuanbin Cheng and Zhentong Shao and Nanpeng Yu and Yu Fu and Yue Dong},
  title     = {Learning to Compress: Energy-Aware Large Language Model Serving via Adaptive {KV} Caching},
  booktitle = {Proc. IEEE Power \& Energy Society General Meeting (PESGM)},
  address   = {Montr\'{e}al, QC, Canada},
  year      = {2026},
  doi       = {10.1109/PESGM58988.2026.11693290}
}

@inproceedings{wan2026grid,
  author    = {Haoxiang Wan and Linhan Fang and Xingpeng Li},
  title     = {Grid Operational Benefit Analysis of Data Center Spatial Flexibility: Congestion Relief, Renewable Energy Curtailment Reduction, and Cost Saving},
  booktitle = {Proc. IEEE Power \& Energy Society General Meeting (PESGM)},
  address   = {Montr\'{e}al, QC, Canada},
  year      = {2026}
}

@article{chung2026openg2g,
  author  = {Jae-Won Chung and Zhirui Liang and Yanyong Mao and Jiasi Chen and Mosharaf Chowdhury and Vladimir Dvorkin},
  title   = {{OpenG2G}: A Simulation Platform for {AI} Datacenter-Grid Runtime Coordination},
  journal = {arXiv preprint arXiv:2605.05519},
  year    = {2026}
}

\end{document}